\documentclass[aip,jcp,,preprint,letterpaper]{revtex4}
\usepackage{comment}
\usepackage{multirow}
\usepackage{array}
\usepackage{bm}
\usepackage{amsmath}
\usepackage{commath}
\usepackage{longtable}
\usepackage{color}
\usepackage{ textcomp }
\usepackage{diagbox}
\usepackage[utf8]{inputenc}
\usepackage{bm}
\usepackage{graphicx}
\usepackage{tikz}
\usetikzlibrary{quantikz}
\usepackage{braket}
\usepackage{float}

\newcommand{\w}{\omega}

\newcommand{\be}{\begin{equation}}
\newcommand{\ee}{\end{equation}}

\begin{document}

\title{Simulation of Condensed-Phase Spectroscopy with Near-term Digital Quantum Computer}
\author{Chee-Kong Lee}
\affiliation{Tencent America, Palo Alto, CA 94306, United States}
\author{Chang-Yu Hsieh}
\affiliation{Tencent, Shenzhen, Guangdong 518057, China}
\author{Shengyu Zhang}
\affiliation{Tencent, Shenzhen, Guangdong 518057, China}
\author{Liang Shi}
\email{lshi4@ucmerced.edu}
\affiliation{Chemistry and Biochemistry, University of California, Merced, California 95343, United States}

\begin{abstract}
Spectroscopy is an indispensable tool in understanding the structures and dynamics of  molecular systems. However computational modelling of spectroscopy is challenging due to the exponential scaling of computational complexity with system sizes unless drastic approximations are made. Quantum computer could potentially overcome these classically intractable computational tasks, but existing approaches using quantum computers to simulate spectroscopy can only handle isolated and static molecules. In this work we develop a workflow that combines multi-scale modeling and time-dependent variational quantum algorithm to compute the linear spectroscopy of systems interacting with their condensed-phase environment via the relevant time correlation function. We demonstrate the feasibility of our approach by numerically simulating the UV-Vis absorption spectra of organic semiconductors. We show that our dynamical approach captures several spectral features that are otherwise overlooked by static methods. Our method can be directly used for other linear condensed-phase spectroscopy and could potentially be extended to nonlinear multi-dimensional spectroscopy. 
\end{abstract}

\clearpage

\maketitle

\section{Introduction}

Condensed phases including liquids and solids host many complex physical,  chemical, and biological processes, such as energy transfer in light-harvesting systems, chemical reactions in liquids, and protein folding in water. Spectroscopy plays a vital role in advancing our understanding of the structure and dynamics of condensed-phase molecular systems. For instance, energy transfer in light-harvesting systems has been monitored by electronic spectroscopy,\cite{Engel2007,Ishizaki2012} while protein structures and their conformational changes can be probed by vibrational spectroscopy.\cite{Ganim2008,Ghosh2017} Although spectroscopy encodes rich molecular information, their interpretations can be challenging and often seek the help from theoretical modeling. 

Spectral modeling is particularly useful for understanding condensed-phase spectroscopy in that many distinct physical phenomena can lead  to similar spectral features. For instance, a broad UV-Vis absorption spectrum of chromophores in solvent may be caused by solvent-induced inhomogeneous broadening, vibronic couplings, or electronic couplings. Because of these possible factors, static (i.e. time-independent) approaches, such as broadening stick spectra from an ensemble of configurations (sometimes called the ensemble approach),\cite{Loco2019a,Zuehlsdorff2019} are often insufficient. Furthermore, the simulation of nonlinear spectroscopy, e.g., pump-probe spectroscopy, often requires the explicit inclusion of the time evolution of the system. Dynamical simulations based on linear and nonlinear response theories have been popular in simulating condensed-phase spectroscopy,\cite{mukamel95,Kowalewski2017,Jansen2019} and the Hamiltonian governing the evolution of the excitations on the chromophores fluctuates due to their environment, sometimes termed the bath. 

However, these dynamical simulations are largely limited to relatively small systems and restricted Hilbert space (e.g., only including single-excitation manifold) due to their high computational cost.\cite{Cho2008,Abramavicius2009} There are scenarios where a larger Hilbert space is needed, e.g., state mixing between ground and low-lying excited states,\cite{Levine2006,Parrish2019} and multiple excitons in light-harvesting systems.\cite{Spano1991,Renger1997,Bruggemann2001,May2014} With the full Hilbert space treatment where all the higher order excitations are also included, the simulation complexity grows exponentially with the number of chromophores. 
This poses great challenges to spectral simulations on current classical computers, and quantum computers offer a potential solution to tackle these classically intractable computational tasks.      

\begin{figure}[ht!]
  \includegraphics[width=1.0\textwidth]{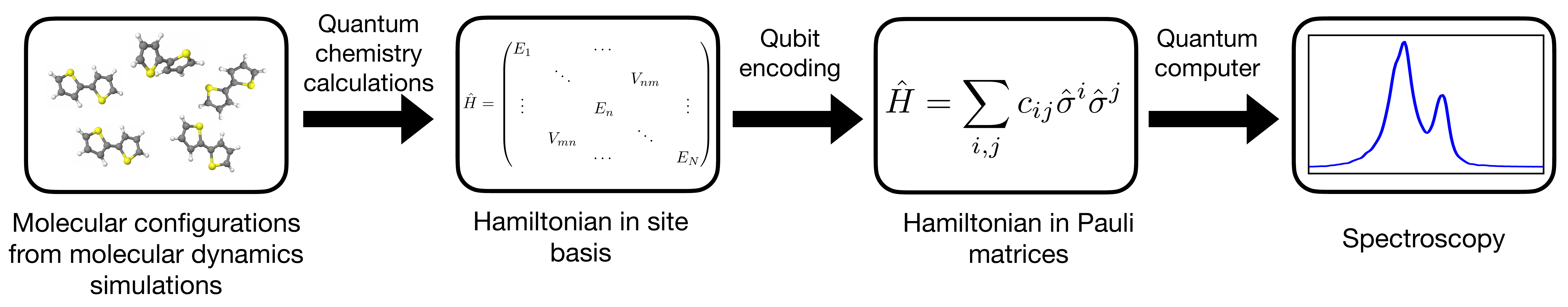}
    \caption{A schematic of the setup in this work. We first perform molecular dynamics (MD) simulations to obtain the molecular configurations. From these molecular configurations, quantum chemistry calculations are utilized to obtain the molecular Hamiltonian in site basis. We then encode the molecular Hamiltonian in terms of Pauli matrices.  Finally we use quantum computer to compute the time correlation function and spectrum. }
    \label{fig:schematic}
\end{figure}

Previous works on quantum simulations of spectroscopy include calculations of vibronic spectra using bosonic sampling, first proposed by Huh and co-workers.\cite{Huh2015} 
The algorithm utilizes the bosonic degrees of freedom in quantum devices to model the vibrational modes in molecules. 
This algorithm has subsequently been experimentally demonstrated in optical,\cite{Clements2018} trapped-ion~\cite{Shen2018} and super-conducting quantum devices.\cite{Wang2020} Recently the algorithm has been extended to include non-Condon effects.\cite{Jnane2020} 
Other than bosonic sampling, calculations of vibronic spectra with digital quantum computers have been proposed by mapping the vibronic degrees of freedom to qubits with binary encoding.\cite{McArdle2019, Sawaya2019}
The calculation of electronic absorption spectroscopy with a multistate contracted variational quantum eigensolver (MC-VQE) has also been proposed for excitonic systems.\cite{Parrish2019} MC-VQE allows efficient computation of the transition energies and oscillator strengths between the ground state and low-lying excited states of a molecule.
All the above approaches are limited to static and isolated systems, thus do not take into account the effects of the condensed-phase environment. 

In this work, we propose calculating condensed-phase spectroscopy with near-term quantum computer and develop a framework that combines multi-scale modeling and variational quantum-classical algorithm. 
Our approach is based on the evaluation of the time-correlation function (TCF) of condensed-phase systems whose Hamiltonians are obtained through molecular dynamics (MD) and electronic structure calculations.
From linear response theory, various types of spectroscopy, e.g., absorption and Raman spectra, can be obtained through the Fourier transform of the TCFs of appropriate operators.
While there are a number of fault-tolerant quantum algorithms for Hamiltonian simulations can be used to compute the TCFs,\cite{Lloyd1996, Childs2012, Low2017, Low2019} these algorithms require long circuit depth and are therefore not suitable for NISQ devices. 
Hence we adopt a time-dependent variational quantum algorithm (VQA) based on the McLachlan's principle and extended it to compute TCF.\cite{Li2017} The VQA was first proposed to simulate the time evolution of wavefunction, and has since been used to simulate dissipative quantum dynamics, linear equations and Green's function.\cite{Endo2019, Endo2020} It has recently been experimentally demonstrated in superconducting quantum devices to simulate adiabatic quantum computing~\cite{Chen2020} and energy transfer.\cite{Lee2021} 

We demonstrate the feasibility of quantum computer in simulating condensed-phase spectroscopy by 
studying the UV-Vis absorption spectrum of bi-thiophene (T2), a widely used organic semiconducting molecules.
The setup of the workflow in this paper is depicted in Fig.~\ref{fig:schematic}. 
We first perform classical MD simulation to obtain the molecular configurations of T2 molecular crystals. For each MD snapshot, quantum chemistry calculations are used to obtain the instantaneous Hamiltonian. 
In particular, we perform time-dependent density function theory (TDDFT) calculations to obtain the excitonic Hamiltonian of multi-chromophoric systems.
Once all the matrix elements of the Hamiltonian along the MD trajectory are computed, the time-dependent Hamiltonian can then be constructed.
The TCF of appropriate operators can then be computed based on the Hamiltonian using the time-dependent VQA and Fourier transformation of TCF is performed to obtain the spectrum. 
Our numerical calculations affirms the feasibility of our approach as we found that the spectra calculated with VQA are in good agreement with numerically exact calculations. Additionally, we compared our approach with the existing static approach used in previous works,\cite{Huh2015,Sawaya2019,Parrish2019} we found that the absorption spectra from VQA exhibit several important spectral features that are not captured by the static approach. The framework developed in this paper can be easily adopted to simulate other linear condensed-phase spectroscopy. It can also be extended to compute the computationally more challenging multi-dimensional spectroscopy.

In the next section, we review the linear response theory and discuss how linear spectroscopy can be obtained from the TCFs. In Sec~\ref{sec:vqa}, we show how the time-dependent VQA can be extended to compute TCFs. We discuss the mapping of the multi-chromophoric excitonic Hamiltonian to Pauli matrices in Sec~\ref{sec:hamiltonian}. The computational details of the Hamiltonian are presented in Sec.~\ref{sec:detail}. In Sec.\ref{sec:numerics}, we show two numerical examples to demonstrate the feasibility of quantum computer to simulate condensed-phase spectra. Finally, we discuss the potential of our method to other related computational problems in Sec.~\ref{sec:discussions}.

\section{Theory of Linear Spectroscopy}
\label{sec:theory}
Within the linear response theory, the line shape of linear spectroscopy, such as absorption and Raman spectra, can be expressed as the Fourier transform of some equilibrium quantum-mechanical TCF,\cite{mcquarrie76}
\be
I(\w) \sim \int_{-\infty}^{\infty} dt\; e^{i\w t} C_{AB}(t), 
\label{eq:ft}
\ee
where the general form of the TCF, $C_{AB}(t)$, for two arbitrary Hermitian operators $\hat A$ and $\hat B$, is given by
\be
C_{AB}(t) = \textrm{Tr} \{ e^{-\beta \hat H_{tot}} e^{i\hat H_{tot} t} \hat B e^{-i \hat H_{tot} t} \hat A \} / \textrm{Tr} \{ e^{-\beta \hat H_{tot}} \},
\label{eq:tcf}
\ee
where $\beta = 1/k_B T$, $\hat H_{tot}$ is the total electronic and nuclear Hamiltonian, and the atomic units are used throughout ($\hbar=1$). For condensed phases, it is customary to take advantage of the separation in the time scales of different degrees of freedom (e.g., electronic versus nuclear motions) and invoke the system-bath treatment, where the system consists of the fast degrees of freedom of interest (e.g., electronic states) with the rest treated as the bath (e.g., nuclear motion). If the energy gaps between ground and excited states of the system are much greater than thermal energy (i.e., $k_B T$), only the ground state will be significantly populated. These approximations lead to the well-known semiclassical expression for TCF, 
\be
C_{AB}(t) = \textrm{Tr}_b \{ \hat \rho_b \bra{G} \hat B(t) \hat U(t)  \hat A(0) \ket{G} \},
\label{eq:aub}
\ee
where $\ket{G}$ is the ground state of the system, $\hat \rho_b = e^{-\beta \hat H_b} /\textrm{Tr}_b \{e^{-\beta \hat H_b}\}$ is the equilibrium ground state density operator for the bath with $\hat H_b = \bra{G} \hat H_{tot} \ket{G}$, and $\textrm{Tr}_b$ indicates a trace over the bath states. The time-dependent operators are defined with respect to the reference ground-state Hamiltonian $\hat H_b$: $ \hat B(t) = e^{i \hat H_b t} \hat B e^{-i \hat H_b t}$, and $\hat U(t)$ is given by
\be
\hat U(t) = \exp_+\left[ -i \int_0^t d\tau \hat{H}(\tau) \right ],
\label{eq:u}
\ee
where $\exp_+$ is the time-ordered exponential, and $\hat{H}(t) = e^{i \hat H_b t} (\hat H_{tot} - \hat H_b) e^{-i \hat H_b t} $. Note that $\hat U(t) \ket{G} = \ket{G}$. For a realistic condensed-phase system, in order to implement Eq. (\ref{eq:aub}), one may approximately treat the bath classically, and its time evolution may be described by classical MD simulation. In doing so, $\hat{B}(t)$ and $\hat{H}(t)$ become classical variables that depends on bath coordinates (e.g., nuclear coordinates).
In some cases when it is inconvenient to implement $\hat A$ and $\hat B$ directly in a quantum circuit, e.g., when they are not unitary, one can use an alternative expression to Eq. (\ref{eq:aub}),   
\be
C_{AB}(t) = \lim_{\lambda \rightarrow 0} \frac{1}{4\lambda^2} \textrm{Tr}_b \{ \hat \rho_b   \bra{G} (e^{-i \lambda  \hat B(t)}-e^{i \lambda \hat B(t)}) \hat U(t) (e^{i \lambda \hat A(0)}-e^{-i \lambda \hat A(0)} ) \ket{G}  \},
\label{eq:aub2}
\ee
where $\lambda$ is a sufficiently small scalar. 

A frequently used approximation to Eq. (\ref{eq:aub}) is to treat the system in the so-called inhomogeneous limit, when the bath correlation time is much longer than the dephasing time, and the bath operators can be considered static, namely, $ \hat A(t)= \hat A(0)$ and $\hat{H}(t)=\hat{H}(0)$. In the inhomogeneous limit, the line shape in the frequency window of interest becomes
\be
I(\omega) \sim \textrm{Tr}_b \left \{ \hat \rho_b \sum_{\alpha} \bra{G} \hat B \ket{\alpha} \delta(\omega - E_{\alpha}) \bra{\alpha} \hat A \ket{G} \right \},
\label{eq:static}
\ee
where $\ket{\alpha}$ is the $\alpha$-th adiabatic excited state, $E_{\alpha}$ is the energy gap between the adiabatic ground and the $\alpha$-th excited states, and $\delta(x)$ is the delta function. The spectrum computed from Eq. (\ref{eq:static}) is often referred to as the static or ensemble spectrum. In Sec.~\ref{sec:numerics} we show that this approach overlooks several spectral features and leads to spectrum that are too broad due to the lack of motional narrowing. 

\section{Computing Correlation Functions with Variational Quantum Algorithm}
\label{sec:vqa} 
In this section we show how near-term quantum computers can be used to compute the TCF of condensed-phase systems. 
We first review the the time-dependent VQA introduced in Ref. \cite{Li2017, Endo2020, Chen2020} in simulating quantum dynamics.
In this VQA, the time-dependent quantum state, $\ket{\Psi(t)}$, is approximated by a parametrized quantum state, $\ket{\psi(\theta(t))}$, i.e.
$\ket{\Psi(t)} =  \hat U(t) \ket{\Psi_0} \approx \ket{\psi(\vec \theta(t))}$, where 
$\vec \theta(t) = [\theta_1(t), \theta_2(t), \theta_3(t) ...] $ denotes the variational parameters at time $t$. According to McLachlan’s principle, the equation of motion for the variational parameters is obtained by minimizing the quantity $\|\Big (i\frac{\partial}{\partial t} -  \hat{H}(t) \Big) \ket{\psi(\vec \theta}\|$ which results in
\begin{eqnarray} \label{eq:EOM_theta}
\vec \theta(t + \delta t) = \vec \theta(t) +  \dot{\vec \theta} (t) \delta t; \,\,\, \dot{\vec \theta} (t) = \hat M^{-1} \vec{V}, 
\end{eqnarray}
where the matrix elements of $\hat M$ and $\vec V$ are
\begin{eqnarray} \label{eq:EOM}
\hat M_{kl} =  \mbox{Re} \left\langle \frac{\partial \psi(\vec{\theta})}{\partial \theta_k} \right\vert \left. \frac{\partial \psi(\vec{\theta})}{\partial \theta_l}
\right\rangle; 
\vec V_{k} =  \mbox{Im} \left\langle  \psi(\vec{\theta}) \right\vert \hat{\tilde{H}} \left\vert \frac{\partial \psi(\vec{\theta})} {\partial \theta_k }
\right\rangle.  
\end{eqnarray}
As described in Ref.\cite{Li2017, Endo2020} these matrix elements can be efficiently obtained in quantum circuits.

Next we discuss how to compute the TCF in Eq. (\ref{eq:aub}). Assuming the operators $\hat A$ and $\hat B$ can be written as a linear combination of Pauli operators (as is the case for dipole operator) $\hat A =\sum_j a_j \hat A_j $ and $\hat B =\sum_i b_j \hat B_i $, the problem of obtaining $C(t)$ is then reduced to computing $\bra{G}\hat B_i \hat U(t) \hat A_j \ket{G}$. 
A straightforward approach to obtain $\bra{G} \hat B_i \hat U(t) \hat A_j \ket{G}$ in a quantum circuit is splitting the evolution operator $\hat U(t)$ via the Lie-Trotter formula. Unfortunately this method requires long circuit depth and quantum error correction which remains elusive in near-term quantum devices. Hence we focus on the hybrid quantum-classical variational algorithm.

\begin{figure}[t!]
  \includegraphics[width=6in]{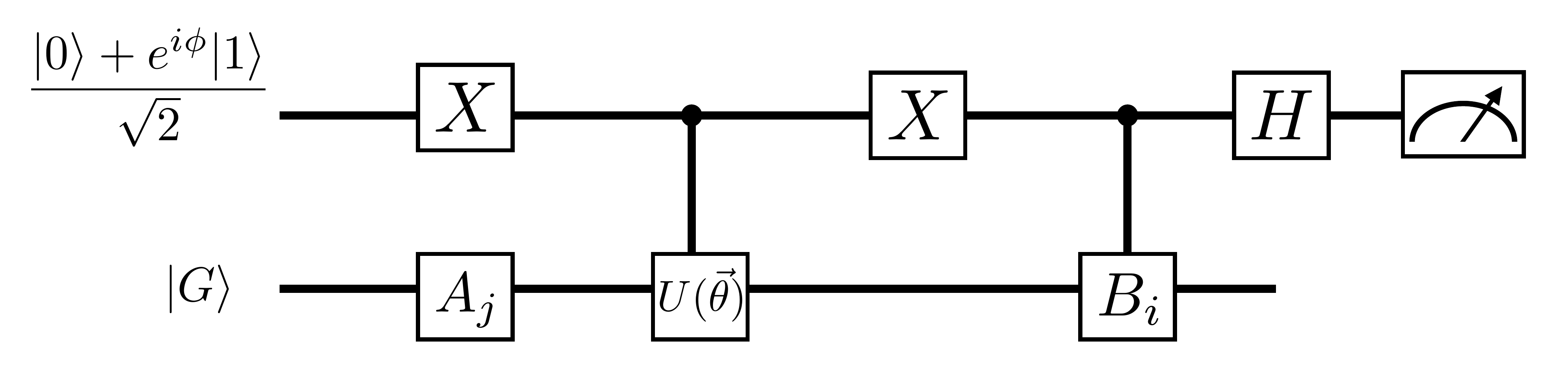}
    \caption{Quantum circuits to compute the transition amplitude, $\bra{G}B_i U(t) A_j \ket{G}$. The ancillary qubit is initialized in state $\frac{\ket{0} + e^{i\phi} \ket{1} }{\sqrt{2}}$. The phase factor $\phi$ is set to be $0$ or $\pi/2$ in order to measure the real and imaginary components of the transition amplitude, respectively.}
    \label{fig:circuit}
\end{figure}

To compute $\bra{G} \hat B_i \hat U(t) \hat A_j \ket{G}$, we first apply $\hat A_j$ to the ground state $\ket{G}$, and then compute the time evolution of the state $\hat A_j \ket{G}$
\begin{eqnarray}
    \mbox{e}_{+}^{-i \int_0^{t'} \hat H(t') dt' }  \hat A_j \ket{G} \approx \hat U(\theta) \hat A_j \ket{G}
\end{eqnarray}
according to Eq. (\ref{eq:EOM_theta}) and (\ref{eq:EOM}) where $\hat U(\theta)$ is parametrized unitary operations in quantum circuits. 
Once the time evolution of the state $\hat A_j \ket{G}$ is obtained, the transition amplitude, $\bra{G}\hat B_i \hat U(t) \hat A_j \ket{G}$, can then be measured using the circuit in Fig.~\ref{fig:circuit}. 
In cases where it is difficult to implement the operator $\hat A_j$ in a quantum circuit, we can prepare the state $\mbox{e}^{i \lambda A_j } \ket{G}$ instead and use the expression in Eq. (\ref{eq:aub2}) to obtain the transition amplitude. 

In the numerical simulations we consider wavefunction ansatz of the form
\begin{eqnarray} \label{eqn:wfn_ansatz}
\ket{\psi(\vec \theta)} = \hat{U}(\vec \theta)\ket{G}  = \prod_{k} \hat{U}_k(\theta_k) \ket{G}  =  \prod_{k} \mbox{e}^{i \theta_k \hat{R}_k} \ket{G}. 
\end{eqnarray}
where $\ket{G}$ is the initial ground state wavefunction and $\hat{R}_k$ is some Pauli strings.
In Sec.~\ref{sec:numerics}, we use an ansatz consisting all combinations of single and 2-qubit rotations. 
Since the Hamiltonians in our numerical examples are real, we drop the terms that only contains single $\hat Y$ terms.
\section{Qubit Hamiltonians for Multichromophoric Systems}
\label{sec:hamiltonian}
For the linear spectroscopy of the systems with $N$ coupled chromophores, one often makes the two-level approximation for each chromophore, and the Hamiltonian can be naturally represented in the diabatic states, $\ket{p_1}\otimes \ket{p_2} \otimes \cdots \otimes \ket{p_N}$, where $p_m$ can be 0 or 1 for $m$th chromophore. In this section, we will present how this diabatic representation of the Hamiltonian for a multichromophoric system can be transformed to that in terms of Pauli matrices (denoted as $I$, $X$, $Y$, $Z$ in this work) under different approximations, a necessary step for quantum computation.

\subsection{Hamiltonian in the Full Hilbert Space}

If the system Hamiltonian only involves one-body and two-body interactions (e.g., electronic Hamiltonian), following the work by Parrish and co-workers,~\cite{Parrish2019} the Hamiltonian in the full Hilbert space $\ket{p_1}\otimes \ket{p_2} \otimes \cdots \otimes \ket{p_N}$ is given by 
\begin{eqnarray} 
\hat{H}(t) = \mathcal{E}(t) \hat I + \sum_{m=1}^N [ \mathcal{Z}_m(t) \hat Z_m + \mathcal{X}_m(t) \hat X_m ] + \sum_{m=1}^N \sum_{n<m} [\mathcal{XX}_{mn}(t) \hat X_m \otimes \hat X_n + \\
\mathcal{XZ}_{mn}(t) \hat X_m \otimes \hat Z_n + \mathcal{ZX}_{mn}(t) \hat Z_m \otimes \hat X_n + \mathcal{ZZ}_{mn}(t) \hat Z_m \otimes \hat Z_n ], \nonumber
\label{eq:H_martinez}
\end{eqnarray}
where the system Hamiltonian matrix elements $\{\mathcal{E}(t),\mathcal{Z}_m(t),\mathcal{X}_m(t),\mathcal{XX}_{mn}(t), \mathcal{XZ}_{mn}(t),\\ \mathcal{ZX}_{mn}(t), \mathcal{ZZ}_{mn}(t)\}$ are time dependent due to the nuclear motion, and their expressions are given in the SI. Here each two-level chromophore is represented by a qubit. It is worthwhile to point out that although in many cases a restricted Hilbert space where only single excitations are allowed may be sufficient, the full Hilbert space is sometimes needed,\cite{Parrish2019} in particular when there are state mixing between the ground and low-lying excited states,\cite{Levine2006} or multiple excitons.\cite{Spano1991,Renger1997,Bruggemann2001,May2014} Simulating the dynamics in the full Hilbert space renders the classical simulation exponentially expensive with the number of chromophores.  

\subsection{Frenkel Exciton Hamiltonian}

In most spectral simulations of multichromophoric systems on classical computers, the Hilbert space is restricted to the single excitations, where only a single $p_m$ is 1 in $\ket{p_1}\otimes \ket{p_2} \otimes \cdots \otimes \ket{p_N}$, and for simplicity we denote these single-excitation diabatic state as $\ket{m}$, meaning that only the $m$th chromophore is on its excited state. If one further assumes the diabatic ground state is decoupled from single-excitation states, the Hamiltonian reduces to the famous Frenkel exciton Hamiltonian,
\be
\hat{H}(t) = \sum_{m=1}^N E_m (t) \ket{m}\bra{m} + \sum_{m=1}^N \sum_{n \ne  m} V_{mn}(t) \ket{m}\bra{n}, 
\label{eq:H_frenkel}
\ee
where $E_m (t)$ is the excitation energy of the $m$th chromophore, and $V_{mn}(t)$ is the (diabatic) couplings between local excitations on the $m$th and $n$th chromophores. To express the Frenkel exciton Hamiltonian in Pauli matrices, we use a standard binary encoding scheme in which the single excitation states of $N$ chromophores can be encoded in the quantum states of  $L=\log_2(N)$ qubits. For example, the state $\ket{m}$ can be represented by
\begin{eqnarray} \label{eqn:state_mapping}
\ket{m} = \ket{\vec{x}} = \ket{x_1} \otimes \ket{x_{2}} \otimes. . . \ket{x_{L}},
\end{eqnarray}
where the subscript denotes the qubit number, $m = x_1 2^0 + x_2 2^1 ... +x_{L}2^{L-1}$ and $x_i$ can be $0$ or $1$. The operator, $\ket{m}\bra{n}$, can be mapped to the qubit representation
\begin{eqnarray} 
\label{eqn:operator_mapping}
\ket{m}\bra{n} = \ket{\vec{x}}\bra{\vec{x}'} =  \ket{x_1}\bra{x'_{1}} \otimes \ket{x_2}\bra{x'_{2}} \otimes ... \otimes\ket{x_{L}}\bra{x'_{L}}, 
\end{eqnarray}
where each binary projector can be expressed in terms of Pauli matrices as follows
\begin{eqnarray}
\ket{0}\bra{1} = \frac{1}{2}(\hat X +  i \hat Y )&;&
\ket{1}\bra{0} = \frac{1}{2}(\hat X -  i \hat Y ); \nonumber \\
\ket{0}\bra{0} = \frac{1}{2}(\hat I +  \hat Z )&;& 
\ket{1}\bra{1} = \frac{1}{2}(\hat I -  \hat Z ).
\end{eqnarray}
With the binary encoding scheme, the Frenkel exciton Hamiltonian of $N$ chromophores leads to a Hamiltonian of $L=\log_2(N)$ interacting qubits, and molecular systems with millions of chromophores (e.g., excitons in organic semiconducting materials) can in principle be studied with as few as tens of qubits, offering the opportunities of simulating quantum dynamics at the nanoscale and its spectroscopic signals quantum mechanically.


\section{Computational Details}
\label{sec:detail}
\begin{figure}[t!]
  \includegraphics[width=6in]{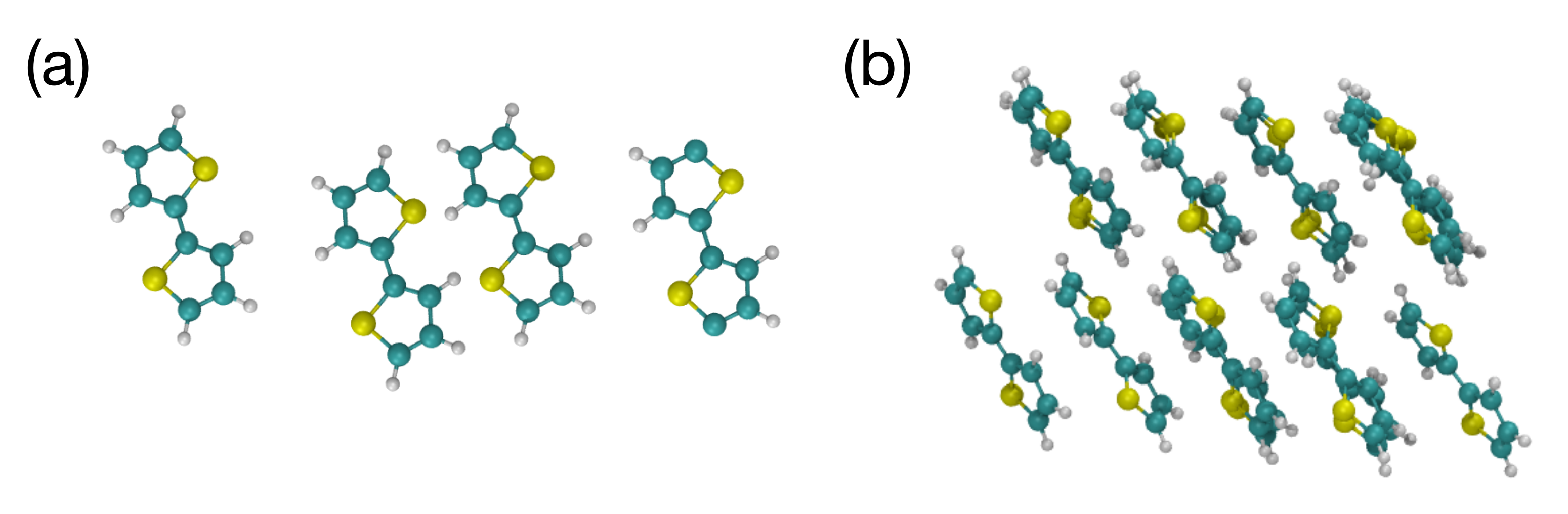}
    \caption{Molecular configurations for the numerical examples in (a) Sec.~\ref{sec:numerics_fci} and (b) Sec.~\ref{sec:numerics_frenkel}. }
    \label{fig:config}
\end{figure}
In this work, we will focus on the ultraviolet-visible (UV-Vis) absorption spectroscopy of organic semiconductors as an example, although the methodology is applicable to other linear spectroscopy, such as electronic circular dichroism and Raman spectroscopy. For UV-Vis, each chromophore can be well approximated as a two-level electronic system, and the bath comprises nuclear degrees of motion. To demonstrate the effectiveness of VQA for realistic molecular systems, classical MD simulation was used to describe the nuclear motion on the electronic ground-state potential energy surface (i.e., $\hat H_b$); hence, the time-dependent operators defined with respect to $\hat H_b$ become classical dynamic variables, and quantum-chemical calculations are performed to obtain their values along the MD trajectory. 

The specific organic semiconducting molecule is bi-thiophene (T2). 
T2 represents the minimum model of oligo- and polythiophenes and their derivatives, one of the most studied semiconducting organic materials.
We use model clusters of 4 and 15 T2 for the Hamiltonian in the full Hilbert space and Frenkel exciton Hamiltonian, respectively, as shown in Fig. ~\ref{fig:config}. These clusters were extracted from the classical MD simulation of a T2 crystal with 32 T2 molecules in the simulation box. The initial simulation box was set up as a $2\times4\times2$ super cell based on the experimental~\cite{Pelletier1994} crystal structure of T2 (CSD identifier: DTENYL02). OPLS/2005 force field was employed in the simulation as it can reasonably reproduce the torsional potential energy surface of T2 predicted from the localized second order M{\o}ller-Plesset perturbation theory (LMP2).\cite{Dubay2012} The MD simulation was performed with the Desmond package 3.6~\cite{Bowers2006} in the NVT ensemble at 133 K, the crystal temperature in the experiment.\cite{Pelletier1994} Temperature was maintained by the Nos\'{e}-Hoover thermostat with a coupling constant of 2.0 ps. Periodic boundary condition was applied to the monoclinic simulation box, and the particle-mesh Ewald (PME) method was employed for electrostatic interactions. The simulation time step was 1 fs, and the configurations were saved every 2 fs during the 10-ps production run for cluster extractions. 

The local excitation energies in the Frenkel exciton Hamiltonian, $E_m$, and the one-body interaction matrix elements for the Hamiltonian in the full Hilbert space (see the SI for details) can be approximately obtained from the lowest-lying excited-state energy of each T2 in the system, and we used TDDFT with the Tamm-Dancoff approximation (TDA) at the theory level of CAM-B3LYP/6-31+G(d). We chose CAM-B3LYP/6-31+G(d) because our previous work~\cite{Lu2020} shows that it gave a good agreement with the result from a correlated wavefunction method (i.e., CC2 method). For the couplings in the Frenkel exciton Hamiltonian, $V_{mn}$, and the two-body interaction matrix elements for the Hamiltonian in the full Hilbert space (see the SI for details), we adopted the dipole-dipole interaction approximation, and the ground-state, excited-state, and transition dipoles for each T2 were computed with DFT or TDDFT at the same theory level. All the quantum chemical calculations were performed with the PySCF program,\cite{Sun2018} and density fitting was used with the heavy-aug-cc-pvdz-jkfit auxiliary basis set implemented in PySCF.

For the calculation of UV-Vis absorption spectra, we also need to construct electric dipole operator, which will substitute the operators $A$ and $B$ in Section \ref{sec:theory}. Since it is a one-body operator, the dipole operator in the full Hilbert space is given by 
\be
\hat \mu^k(t) = \sum_m \hat \mu^k_m(t) =  \sum_m [  \mu_{I,m}^k(t) \hat I_m + \mu_{Z,m}^k(t) \hat Z_m + \mu_{X,m}^k(t) \hat X_m ] ,
\label{eq:mu_martinez}
\ee
where $\hat \mu^k_m$ is the $k$th Cartesian component of the dipole operator ($k=x,y,z$) for the $m$th T2 molecule, and the expressions for $\mu_{I,m}^k$, $\mu_{Z,m}^k$, and $\mu_{X,m}^k$ are given in the SI. It is worth noting that the normalized initial state $\hat \mu_m^k \ket{G}/\norm{\hat \mu_m^k}$ can be efficiently prepared since it only involves single-qubit operations.
The single qubit rotations, $\hat R$, for preparing such  initial state can be computed classically via the relation $  \hat R(\phi)\ket{G} =\hat  \mu_m^k \ket{G}/\norm{\hat \mu_m^k}$. 

In the restricted Hilbert space for the Frenkel exciton Hamiltonian, the dipole operator is given by
\be
\hat \mu^k(t) =  \sum_m \hat \mu^k_m(t) = \sum_m \mu_{m}^k(t) (\ket{G} \bra{m} + \ket{m} \bra{G}),
\label{eq:mu_frenkel}
\ee
where $\mu_{m}^k$ is the $k$th Cartesian component of the transition dipole associated with the excitation localized on the $m$th T2 molecule, and $\ket{G} \bra{m}$ and $\ket{m} \bra{G}$ will be mapped to Pauli matrices using the binary encoding scheme. After the encoding the dipole operator could consist of many-body qubit interaction terms, and this makes the preparation of the initial state difficult since $\hat \mu^k_m$ is non-unitary. Thus we use Eq. (\ref{eq:aub2}) for the evaluation of TCF within the Frenkel exciton basis and we use $\lambda=0.1$.
All the time-dependent coefficients in Eqs. (\ref{eq:mu_martinez}) and (\ref{eq:mu_frenkel}) can be obtained from the ground-state, excited-state, and transition dipoles of T2 molecules over the MD trajectory. 

In using Eq. (\ref{eq:ft}) to compute the absorption spectra, we applied an exponential damping function to reduce spurious effects in the Fourier transform,\cite{Valleau2012,Loco2019a} namely
\be
I(\w) \sim \int_{-\infty}^{\infty} dt\; e^{i\w t} C_{\mu\mu}(t) e^{-|t|/\tau} , 
\label{eq:ft_lt}
\ee
where $C_{\mu\mu}(t)$ is the dipole-dipole TCF for absorption spectrum, and $\tau$ is the decay constant. In some cases, this damping function can also be used to phenomenologically include lifetime effects, for which $\tau = 2 T_1$ and $T_1$ is the lifetime. In the inhomogeneous limit, applying this damping function is equivalent to broadening the static spectrum, Eq. (\ref{eq:static}), with a Lorentzian, leading to the following expression for absorption line shape
\be
I(\omega) \sim \textrm{Tr}_b \left \{ \hat \rho_b \sum_{\alpha} |\bra{G} \hat \mu \ket{\alpha}|^2 \frac{1/\tau}{(\omega-E_{\alpha})^2 + (1/\tau)^2} \right \},
\label{eq:static_lt}
\ee
where the summation is over eigenstates of $\hat H$. In this work, we take $\tau=50$ fs, and we show in the SI that the specific (reasonable) value of $\tau$ has little effect on the spectrum computed from TCF, but affects the static spectrum from Eq. (\ref{eq:static_lt}) significantly. For simplicity but without losing generality, we took the average dipoles in our spectral simulations (since dipole fluctuation in a crystal is very small), and assumed that the T2 clusters are oriented isotropically such that the dipole-dipole TCF is simply the average of the three Cartesian components, namely,
\be
C_{\mu\mu}(t) = \frac{1}{3}\left[ C_{\mu_x\mu_x}(t) + C_{\mu_y\mu_y}(t) + C_{\mu_z\mu_z}(t)\right].
\ee
In our numerical simulations, we take average TCFs of 400 MD trajectories. 


\section{Numerical Results and Discussions}
\label{sec:numerics}
\subsection{Ab Initio Exciton Hamiltonian in the Full Hilbert Space}
\label{sec:numerics_fci}

\begin{figure}[ht!]
  \includegraphics[width=4.5in]{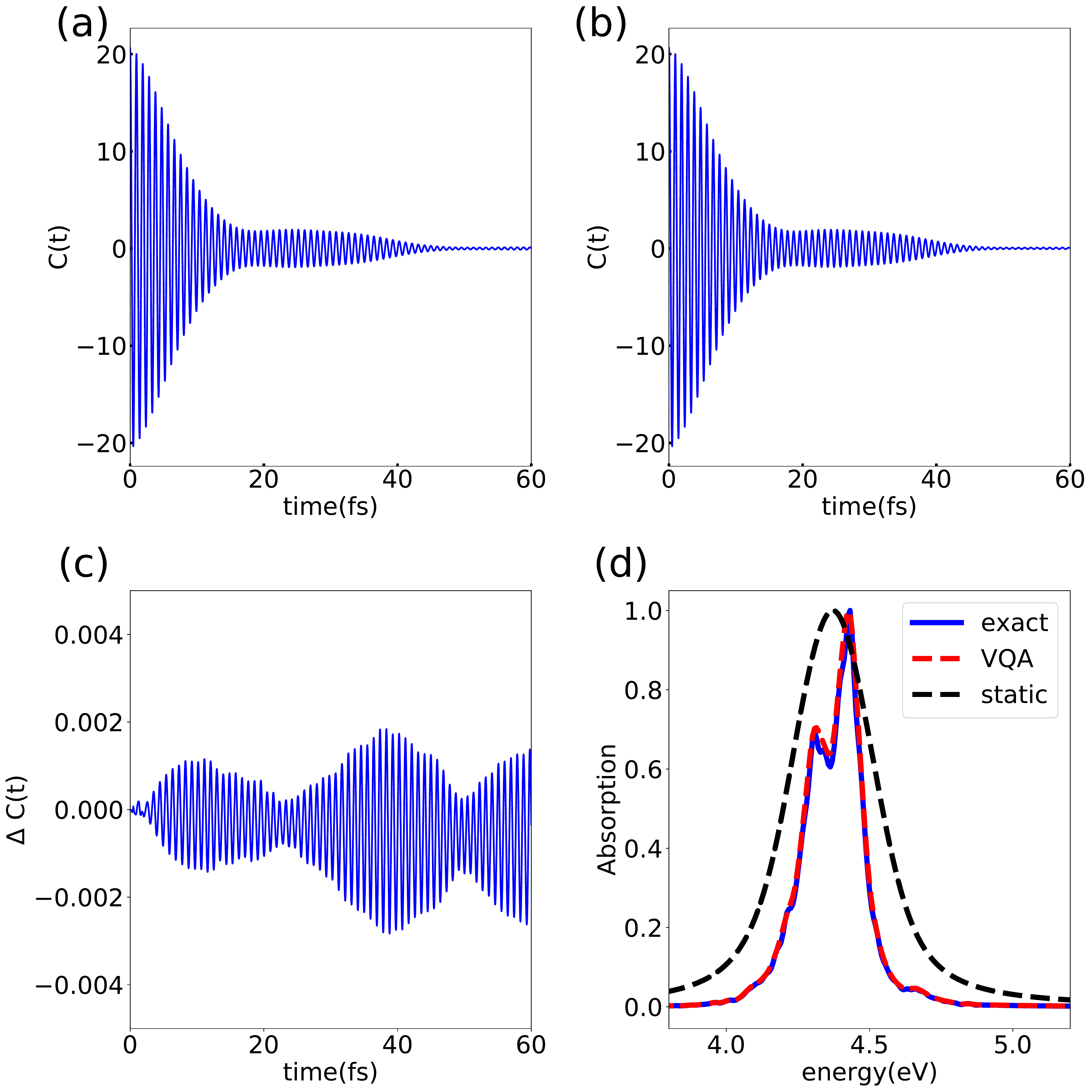}
    \caption{Numerical results with the full Hilbert space treatment. (a) Dipole-dipole TCF from numerically exact calculations. (b) Dipole-dipole TCF from VQA. (c) Relative difference between the TCFs from exact calculations and VQA. (d) Absorption spectra obtained from Fourier transform of TCFs along with the static spectrum computed from Eq. (\ref{eq:static_lt}).}
    \label{fig:fci}
\end{figure}
Here we numerically simulate the absorption spectrum of a system consisting of 4 T2 molecules, and its ab initio exciton Hamiltonian in the full Hilbert space is given by Eq. (\ref{eq:H_martinez}). 
We first consider the dipole-dipole TCFs obtained from numerically exact simulations and VQA, as shown in Fig.~\ref{fig:fci}(a) and (b), respectively. 
The exact TCF exhibits two distinct time scales: a fast oscillation and a slower decay. The fast oscillation with a sub-femtosecond period is a result of the transitions between the ground and excited states (the monomer has a transition energy of about 4 to 5 eV). The slow decay on the time scale of tens of femtosecond results from the coupling to the nuclear motion (i.e., dephasing).
It can be seen from Fig.~\ref{fig:fci} (b) that the TCF computed from VQA is capable of capturing both time scales accurately. 
To gain further insight into the errors incurred in the VQA, we computed the normalized difference between the TCFs from the exact calculation and VQA, $\Delta C(t) = (C_{exact}(t) - C_{VQA}(t))/C_{exact}(0)$, shown in Fig.~\ref{fig:fci} (c). The relative errors of VQA are very small at less than $0.005$. 
It will be worthwhile to explore how robust the VQA approach is against the imperfections of actual quantum devices, such as the state preparation and measurement (SPAM) errors. 

The absorption spectra from the numerically exact and VQA calculations are shown in Fig.~\ref{fig:fci}(d) as blue and red lines, respectively, which are hardly distinguishable. Both methods show a main peak at 4.5 eV and a lower peak at 4.3 eV. This double peak feature might be concealed if one computes the spectra using a static approach with artificial broadening. In Fig.~\ref{fig:fci} the spectrum computed using Eq. (\ref{eq:static_lt}) with the same decay constant of 50 fs is shown as the dashed black line, and only a single broad peak is observed. Furthermore, we show in the SI that the absorption spectrum from the static approach is highly sensitive to the phenomenological decay time constant, $\tau$, in Eq. (\ref{eq:static_lt}), whereas the dynamical calculation of spectrum is largely independent of the choice of the decay time constant. As long as the chosen decay time constant is not significantly shorter than the (pure) dephasing time, the line width of the spectrum is mostly determined by the latter, and the effect of the decay function is merely smoothing the spectrum. For the static approach, due to the lack of motional narrowing, the spectrum is often too broad compared to that from dynamical calculations. This clearly demonstrates the necessity of adopting a dynamical approach to model the absorption spectrum of condensed phases. 

\subsection{Frenkel Exciton Hamiltonian}
\label{sec:numerics_frenkel}

\begin{figure}[ht!]
  \includegraphics[width=4.5in]{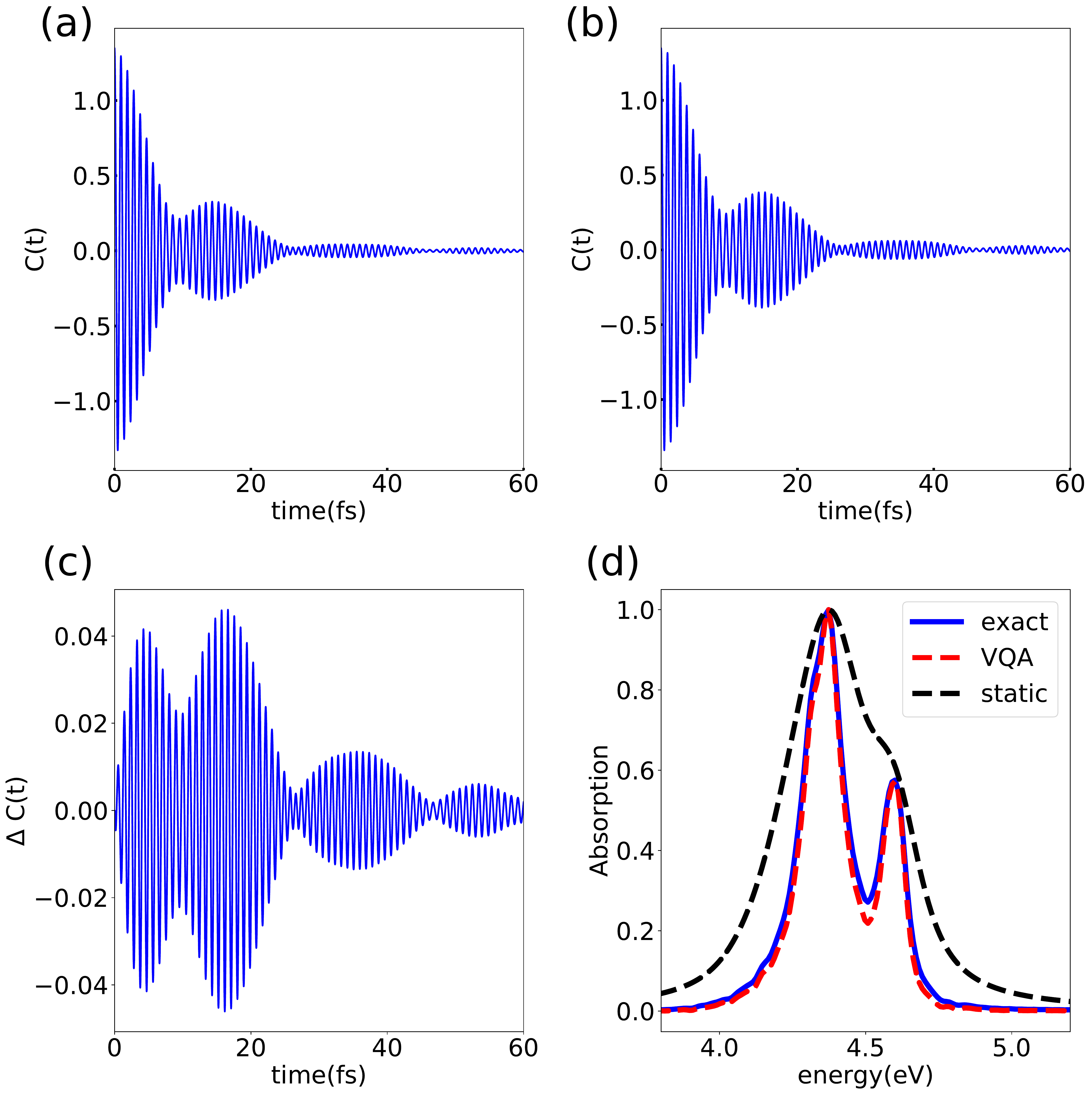}
    \caption{Numerical results with within the Frenkel exciton approximation (i.e. single excition). (a) Dipole-dipole TCF from numerically exact calculations. (b) Dipole-dipole TCF from VQA. (c) Relative difference between the TCFs from exact calculations and VQA. (d) Absorption spectra obtained from Fourier transform of TCFs along with the static spectrum computed from Eq. (\ref{eq:static_lt}).}
    \label{fig:frenkel}
\end{figure}

Next we consider a system of 15 T2 molecules within the Frenkel exciton basis (i.e., single excitation), and the corresponding Hamiltonian is Eq. (\ref{eq:H_frenkel}). 
Using the binary encoding scheme, the system can be encoded into 4 qubits since it contains 16 quantum states (15 excited states and 1 ground state).
Similar to the previous section, we computed the TCFs from numerically exact simulations and VQA, the resulting TCFs are shown in Fig.~\ref{fig:frenkel}(a) and (b). In obtaining the TCF from VQA, Eq. (\ref{eq:aub2}) was used with $\lambda=0.1$, and we verified that smaller $\lambda$ values do not make any discernible changes.
Good agreement is again seen between the VQA and exact results, and the relative error of VQA remains below $0.05$ throughout the time evolution, as shown in Fig.~\ref{fig:frenkel}(c). 
Note that this error, although still small, is greater than that in the previous section, and this is likely due to the more complex Hamiltonian after the binary encoding (e.g., it involves 4-qubit interaction terms).
The resulting absorption spectra after Fourier transform of the TFCs are plotted in Fig.~\ref{fig:frenkel}(d), and the good agreement of the VQA calculation (dashed red line) with exact calculation (solid blue line) further confirms the capability of the VQA in modeling absorption spectroscopy. In comparison, the spectrum from the static approach using Eq. (\ref{eq:static_lt}) (dashed black line) predicts a much broader spectrum due to the absence of motional narrowing, and consequently the peak at $4.6$ eV becomes a barely visible shoulder of the main peak. It is worthwhile to point out that there is no straightforward method to choose a proper broadening parameter (i.e., $\tau$ in Eq. (\ref{eq:static_lt})) a priori.  

\section{Conclusions and Outlook}
\label{sec:discussions}
In recent years there have been a large number of works developing NISQ algorithms to compute the ground or low-energy states of static and isolated quantum systems.\cite{McArdle2020, Cao2019, Bauer2020, Hsieh2021, Peruzzo2014, Takeshita2020, Ollitrault2020, Xia2018, AIQuantum2020}
However, chemical systems are typically immersed in condensed phases and many phenomena like chemical reactions and spectroscopy are heavily influenced by the interactions with these condensed-phase environments. 
Computational modelling of these phenomena is challenging as it requires simulating the time evolution of quantum dynamics, a computational task that is far more difficult than the ground state simulation as quantum dynamics typically involve the participation of many eigenstates. 
The framework developed in this work combining multi-scale modeling and quantum algorithm could potentially overcome these challenges.
Specifically, we have demonstrated how near-term quantum computers can be used to simulate condensed-phase spectroscopy by modeling UV-Vis absorption spectrum as an example. Since the excitation of the chromophores of interest is heavily influenced by their local molecular environment in condensed phases, we employed a dynamical approach based on the linear response theory, and extend the hybrid quantum-classical variational algorithm to compute the relevant TCF on near-term quantum computers. We examined two types of Hamiltonians: one is in the full Hilbert space with one qubit representing each chromophore, and the other is the Frenkel exciton Hamiltonian with $N$ chromophores mapped to $\log_2(N)$ qubits using the binary encoding scheme. The matrix elements in the Hamiltonians were obtained from MD simulations and quantum-chemical calculations to provide a realistic description of the Hamiltonians and their time evolution. Our numerical simulations show that in both cases the variational algorithm can faithfully reproduce the spectra from exact calculations. We also compared the spectra from our dynamical approach to those from the static ensemble method, and the comparison clearly shows the necessity of adopting the dynamical approach to simulate condensed-phase spectroscopy.

Despite the success of VQA in numerical spectral simulation, it would be imperative to experiment VQA on actual near-term quantum devices to examine its robustness against device imperfection. A recent work by some of us shows that error mitigation is essential in the actual implementation of VQA on NISQ devices for the dynamical simulation of energy transfer process, so we anticipate the same situation for spectral simulation. As our approach is based on the computation of TCFs, we envision that VQA may be applicable to other problems where TCF is the key quantity, such as quantum transport with a full Hilbert space treatment. The power of the dynamical approach to spectral simulation is presumably more manifested in the simulation of nonlinear spectroscopy, and it would be interesting to extend current method to model them in the future. An alternative route to the modeling of condensed-phase spectroscopy is using the methods for open-quantum systems with density matrices (e.g., quantum master equation), which may overcome some shortcomings of the current semiclassical method, e.g., classical treatment of the bath. Several quantum algorithms have been proposed to simulate the dynamics of open quantum systems~\cite{Yuan2019, Hu2020, Lee2020a}, and it will be a worthwhile endeavour to incorporate these algorithms to simulate condensed-phase spectra.

\section{Acknowledgment}
L.S. acknowledges the support from the University of California Merced start-up funding. 

\clearpage
\bibliographystyle{naturemag_noURL}
\bibliography{MyCollection.bib}

\end{document}


\newcommand{\wn}{cm$^{-1}$}
\newcommand{\td}{$\sim$}
\newcommand{\la}{\langle}
\newcommand{\ra}{\rangle}
\newcommand{\e}{\epsilon}
\newcommand{\w}{\omega}
\newcommand{\bracket}[1]{\left\langle #1 \right\rangle}
\newcommand{\degreec}{^{\circ}{\rm C}}
\newcommand{\be}{\begin{equation}}
\newcommand{\ee}{\end{equation}}
\newcommand{\ie}{{\it i.e.}}
\newcommand{\eg}{{\it e.g.}}
\newcommand{\etal}{{\it et al.}}
\newcommand{\bra}[1]{\left<#1\right|}
\newcommand{\ket}[1]{\left|#1\right>}
\newcommand{\ketbra}[2]{\ket{#1}\bra{#2}}
\renewcommand{\thepage}{S\arabic{page}}

\title{{\huge Supplementary Materials}\\
\vskip 0.2in
Simulation of Condensed-Phase Spectroscopy with Near-term Digital Quantum Computer}

\author{Chee-Kong Lee}
\affiliation{Tencent America, Palo Alto, CA 94306, United States}
\author{Chang-Yu Hsieh}
\affiliation{Tencent, Shenzhen, Guangdong 518057, China}
\author{Shengyu Zhang}
\affiliation{Tencent, Shenzhen, Guangdong 518057, China}
\author{Liang Shi}
\email{lshi4@ucmerced.edu}
\affiliation{Chemistry and Biochemistry, University of California, Merced, California 95343, United States}

\maketitle

\section{Mapping Exciton Hamiltonians to Pauli matrices}
The exciton Hamiltonian in the full Hilbert space of $N$ two-level chromophores can be expressed in terms of Pauli matrices, and the details have been worked out by Parrish and co-workers.\cite{Parrish2019} We follow their notations which are based on the chemists' notations for matrix elements, and only briefly review the results here. Based on the relationships between the binary projectors and the Pauli matrices, shown in Eq. (15) of the main text, the Hamiltonian including one-body and two-body interactions can be written as 
\begin{eqnarray} 
\hat{H}  = \mathcal{E} \hat I + \sum_{m=1}^N ( \mathcal{Z}_m \hat Z_m + \mathcal{X}_m \hat X_m ) + \sum_{m=1}^N \sum_{n<m} (\mathcal{XX}_{mn} \hat X_m \otimes \hat X_n + \\
\mathcal{XZ}_{mn} \hat X_m \otimes \hat Z_n + \mathcal{ZX}_{mn} \hat Z_m \otimes \hat X_n + \mathcal{ZZ}_{mn} \hat Z_m \otimes \hat Z_n ), \nonumber
\label{eq:H_martinez}
\end{eqnarray}
where the matrix elements are given by
\begin{eqnarray} 
\mathcal{E} = \sum_m S_m + \sum_{n<m} (S_m|S_n)\\
\mathcal{Z}_m = D_m + \sum_n (D_m|S_n) \\
\mathcal{X}_m = X_m + \sum_n (T_m|S_n) \\
\mathcal{XX}_{mn} = (T_m|T_n) \\
\mathcal{XZ}_{mn} = (T_m|D_n)\\
\mathcal{ZX}_{mn} = (D_m|T_n)\\ 
\mathcal{ZZ}_{mn} = (D_m|D_n).
\end{eqnarray}
In these expressions, $S_m$, $D_m$, and $X_m$ are determined by the matrix elements of one-body term, $\hat h$, associated with the $m$th chromophore,
\begin{eqnarray} 
S_m = [ ( 0_m | \hat h | 0_m ) + ( 1_m |  \hat h | 1_m ) ] /2 \\ 
D_m = [ ( 0_m | \hat h | 0_m ) - ( 1_m |  \hat h | 1_m ) ] /2\\
X_m = ( 0_m | \hat h | 1_m ),
\end{eqnarray}
where $\ket{0_m}$ and $\ket{1_m}$ are the ground and excited states of the $m$th chromophore. The terms of the form $(\cdots|\cdots)$ are associated with the two-body interaction term, $\hat v$, and symbols $|S_m)$, $|D_m)$, and $|T_m)$ denote
\begin{eqnarray} 
|S_m) = |0_m 0_m + 1_m 1_m )/2 \\ 
|D_m) = |0_m 0_m - 1_m 1_m )/2\\
|T_m) = |0_m 1_m).
\end{eqnarray}
For example, $(T_m | S_n) = (0_m 1_m | \hat v | 0_n 0_n + 1_n 1_n)/2$. Within the Born-Oppenheimer approximations, all the matrix elements are still functions of nuclear coordinates, and their time dependence is governed by the classical MD simulation. In implementing this Hamiltonian for T2 clusters, we approximately set $( 0_m | \hat h | 0_m )=( 0_m | \hat h | 1_m )=0$, and $( 1_m | \hat h | 1_m )$ to be the lowest lying excited state energy of the $m$th T2 molecule from TDDFT calculations. Two-body interaction matrix elements were computed using the approximation of dipole-dipole interaction, and the dipole position is taken to be the centers of mass of the T2 molecules. Ground-state, excited-state and transition dipoles were used for the terms involving $|0_m 0_m)$, $|1_m 1_m)$, and $|0_m 1_m)$, respectively. For example, 
\be
(0_m 0_m| \hat v | 0_n 1_n) \approx \frac{\vec{\mu}_{00,m} \cdot \vec{\mu}_{01,n} - 3(\vec{\mu}_{00,m} \cdot \hat{r}_{mn})(\vec{\mu}_{01,n} \cdot \hat{r}_{mn}) }{r_{mn}^3}, 
\ee
where $\vec{\mu}_{00,m}$ is the ground-state dipole of the $m$th T2, $\vec{\mu}_{01,n}$ is the transition dipole associated with the excitation on the $n$th T2, $\vec{r}_{mn}$ is the vector connecting the centers of mass of the two T2 molecules, $r_{mn} = |\vec{r}_{mn}|$, and $\hat{r}_{mn} = \vec{r}_{mn} / |\vec{r}_{mn}|$. In the same basis of Pauli matrices, the $k$th Cartesian component of the dipole operator is given by
\be
\hat \mu^k(t) = \sum_m [  \mu_{I,m}^k(t) \hat I_m + \mu_{Z,m}^k(t) \hat Z_m + \mu_{X,m}^k(t) \hat X_m ] ,
\label{eq:mu_martinez}
\ee
where the matrix elements are 
\begin{eqnarray} 
\mu_{I,m}^k = (\mu_{00,m}^k + \mu_{11,m}^k)/2 \\ 
\mu_{Z,m}^k = (\mu_{00,m}^k - \mu_{11,m}^k)/2\\
\mu_{X,m}^k = \mu_{01,m}^k,
\end{eqnarray}
where $\mu_{00,m}^k$, $\mu_{11,m}^k$, and $\mu_{01,m}^k$ are the $k$th Cartesian coordinates of the ground-state, excited-state and transition dipoles for the $m$th chromophore, which were obtained from (TD)DFT calculations. 

For the Frenkel exciton Hamiltonian (Eq. (12) in the main text), where only the single excitations are included in the basis, the diagonal matrix elements are the excitation energies of the chromophores computed from TDDFT, and the off-diagonal elements are approximated by transition dipole couplings. The only non-zero matrix elements in the dipole operator are the transition dipoles for each chromophore, i.e., $\vec{\mu}_{01,m}$.

\section{Effects of Exponential Decay Function}


Fig.~\ref{fig:martinez_lifetime} displays the absorption spectra of 4 T2 computed from the static and dynamical approaches, using Eq. (18) and (19) in the main text, respectively, for the excitonic Hamiltonian in the full Hilbert space. It is clear that the spectrum from the dynamical approach is not sensitive to the chosen decay time constant, $\tau$, whereas the static approach is. This observation holds for the spectra of 15 T2 using the Frenkel exciton Hamiltonian, as shown in Fig. \ref{fig:frenkel_lifetime}.

\renewcommand{\thefigure}{S1}
\begin{figure}[h!tbp]
  \centering
    \includegraphics[width=0.6\textwidth]{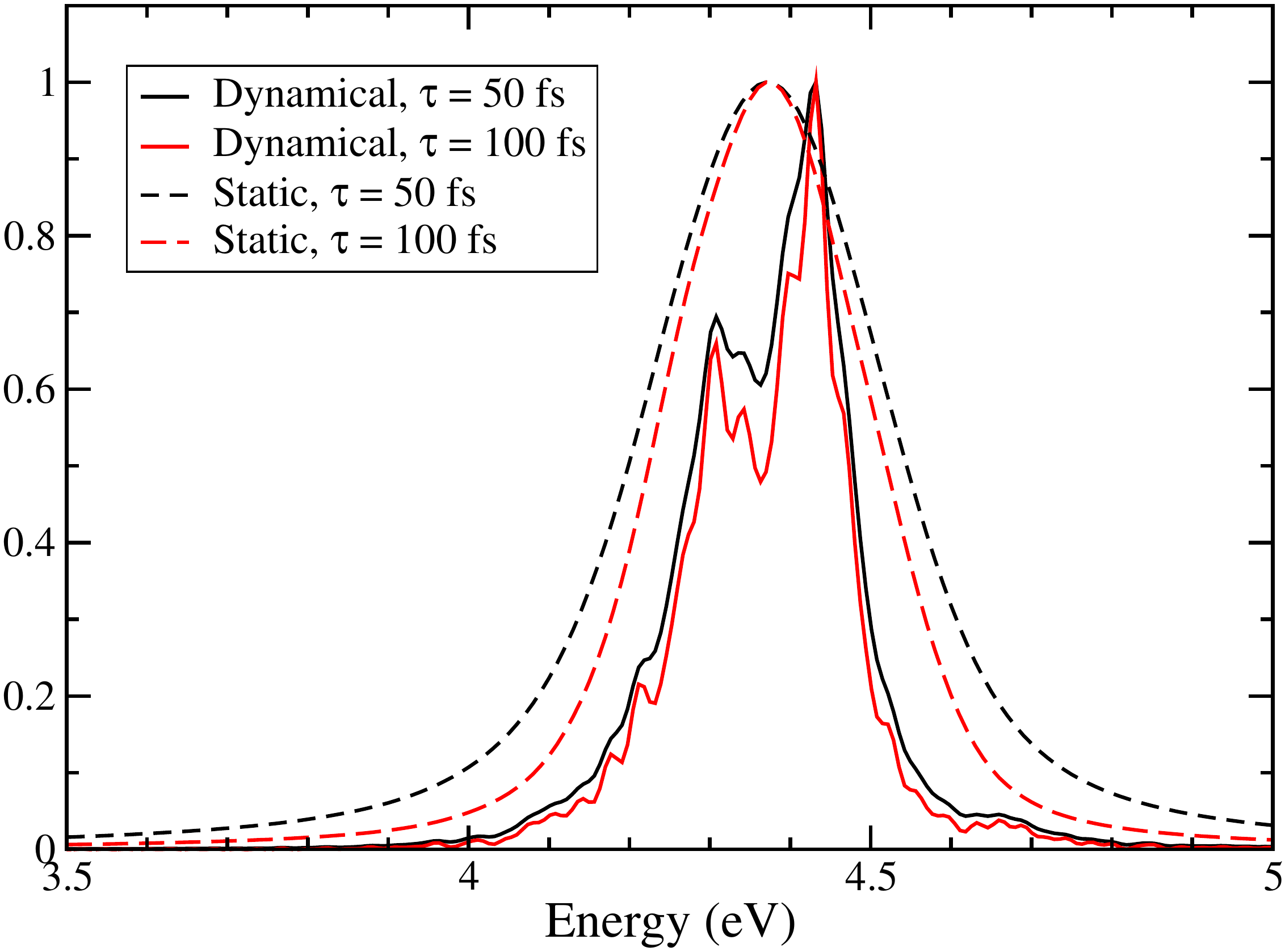}
     \caption{Effects of the exponential decay time constant, $\tau$, on the computed absorption spectra of 4 T2 using static and dynamical approaches for the exciton Hamiltonian in the full Hilbert space.}
     \label{fig:martinez_lifetime}
\end{figure}

\renewcommand{\thefigure}{S2}
\begin{figure}[h!tbp]
  \centering
    \includegraphics[width=0.6\textwidth]{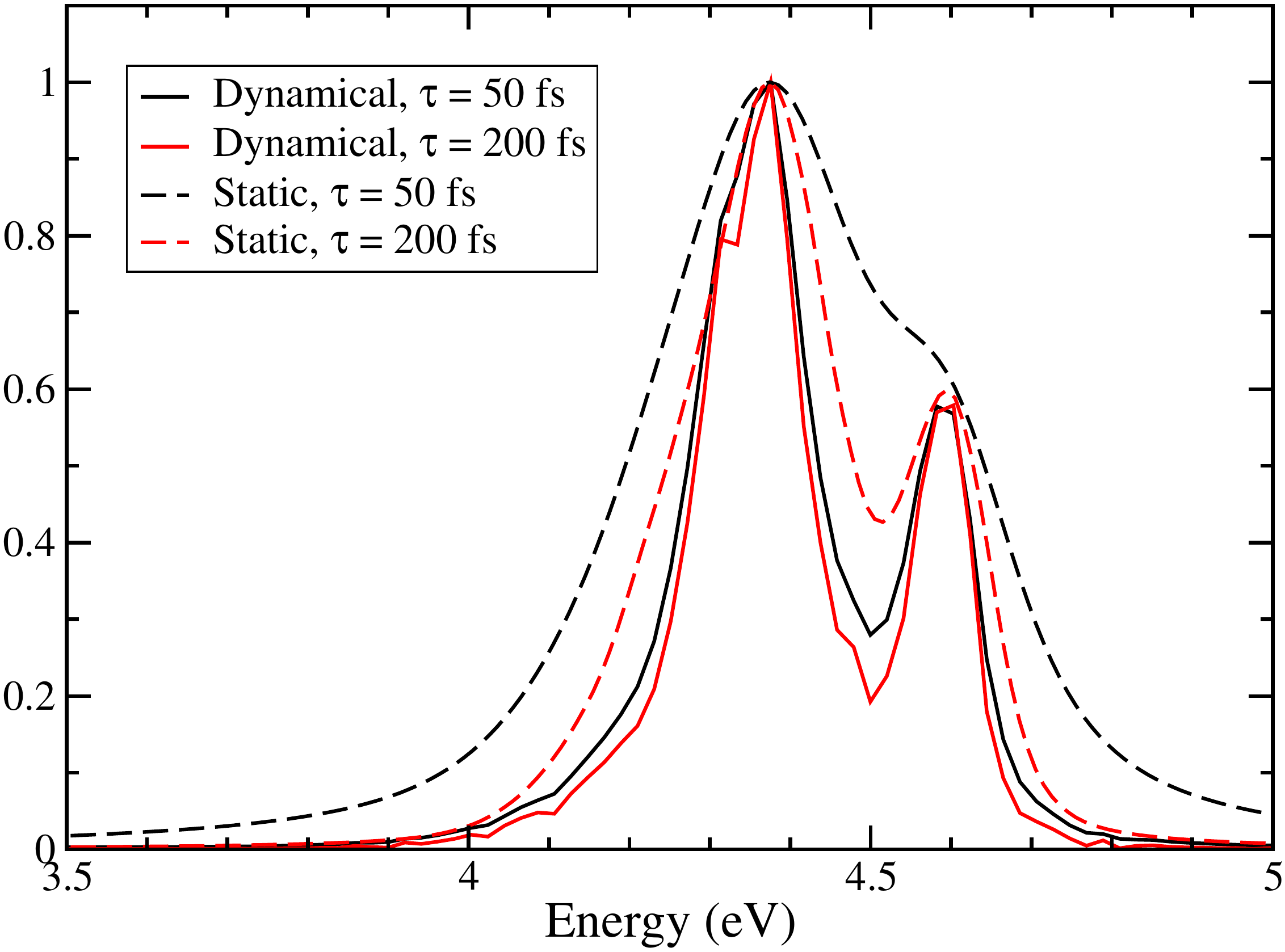}
     \caption{Effects of the exponential decay time constant, $\tau$, on the computed absorption spectra of 15 T2 using static and dynamical approaches for the Frenkel exciton Hamiltonian}
     \label{fig:frenkel_lifetime}
\end{figure}

\clearpage
\bibliographystyle{naturemag_noURL}
\bibliography{MyCollection}